\documentclass[letter]{aa}
\usepackage{graphicx}
\usepackage{txfonts}
\usepackage{natbib}
\usepackage{verbatim}
\usepackage{txfonts,epsfig,graphicx,url,twoopt}
\usepackage[usenames,dvipsnames]{color}
\usepackage[breaklinks=true]{hyperref}
\hypersetup{colorlinks=true,citecolor=blue}
\bibpunct{(}{)}{;}{a}{}{,} 
\usepackage{hyperref}

\begin{document}

   \title{Asymmetric transition disks: Vorticity or eccentricity?}

   \author{S.~Ataiee\inst{1,2}
          \and
          P.~Pinilla\inst{1}
          \and
          A.~Zsom\inst{3}
          \and
          C.P.~Dullemond\inst{1}
          \and
          C.~Dominik\inst{4,5}
          \and
          J.~Ghanbari\inst{2}
          }

   \institute{Heidelberg University,
              Center for Astronomy,
              Institute for Theoretical Astrophysics,
              Albert Ueberle Str.~2, 69120 Heidelberg, Germany \email{sareh.ataiee@gmail.com}
         \and
        			 Department of Physics,
             Faculty of Sciences,
             Ferdowsi University of Mashhad, Mashhad, Iran
         \and
             Department of Earth,
             Atmospheric and Planetary Sciences,
             Massachusetts Institute of Technology, Cambridge, MA 02139, USA
         \and
             Anton Pannekoek Institute for Astronomy,
             University of Amsterdam,
             Postbus 94249, 1090 GE Amsterdam, the Netherlands
         \and
         		 Afdeling Sterrenkunde,
         		 Radboud Universiteit Nijmegen,
         		 Postbus 9010, 6500 GL, Nijmegen, the Netherlands          
             }

   \date{XXXX}

 \abstract{
  Transition disks typically appear in resolved millimeter observations as
   giant dust rings surrounding their young host stars. More accurate observations with ALMA have shown several
   of these rings to be in fact asymmetric: they have lopsided shapes. It has been
   speculated that these rings act as dust traps, which would make them important laboratories for studying planet formation. It has been shown that an elongated giant vortex produced in a disk with a strong viscosity jump strikingly resembles the observed asymmetric rings.  
}{
   We aim to study a
  similar behavior for a disk in which a giant planet is embedded.  However,
  a giant planet can induce two kinds of asymmetries: (1) a giant
  vortex, and (2) an eccentric disk. We studied
  under which conditions each of these can appear, and how one can
  observationally distinguish between them. This is important because only
  a vortex can trap particles both radially and azimuthally, while the
  eccentric ring can only trap particles in radial direction.
}{
  We used the FARGO code to conduct the hydro-simulations. We set up a
  disk with an embedded giant planet and took a radial grid spanning from
  0.1 to 7 times the planet semi-major axis. We ran the simulations with various viscosity values and planet masses for 1000
  planet orbits to allow a fully developed vortex or disk eccentricity. Afterwards, we compared the dust distribution in a vortex-holding disk with an eccentric disk using dust simulations.
}{
 We find that vorticity and eccentricity are distinguishable by looking at the azimuthal contrast of the dust density. While vortices, as particle traps, produce very pronounced azimuthal asymmetries, eccentric features are not able to accumulate millimeter dust particles in azimuthal direction, and therefore the asymmetries are expected to be modest.
}{}


   \maketitle
\section{Introduction} \label{intro}
Most models of protoplanetary disks that aimed at fitting observations are axisymmetric \citep[e.g.][]{D'AlessioPaola1998,Pinte2006,Woitke2009,Min2011} because until recently, with a few exceptions, observations did not have enough spatial resolution and sensitivity to detect strong deviations from axial symmetry in these disks. Recent observations, however, have unveiled some non-axisymmetric structures in protoplanetary disks and have changed the situation. Scattered-light images of a number of disks around Herbig Ae/Be stars have revealed complex structures such as spirals and rings
\citep[e.g.][]{Fukagawa2004,Oppenheimer2008,Muto2012}. Millimeter continuum maps show asymmetries as well \citep{Pietu2005}. Spatially resolved mm maps consistently show that transitional disks -protoplanetary disks with a strong deficiency of dust inward of a few AU, or in other words, a disk with a huge inner hole \citep{Andrews2011}- appear as rings on the sky often with a lumpy structure along the ring \citep{Brown2009,Casassus2013}. Near-infrared scattered-light images of some of these disks also clearly show non-axisymmetric structures \citep{Rameau2012}. Altogether, non-axisymmetry turns out to be a regular phenomenon in disks, particularly in transition disks.

Recent theoretical work has aimed at explaining the origin and physical processes involved in these asymmetries \citep[e.g.][]{Birnstiel2013a}. For instance, \cite{Regaly2011} and \cite{Lyra2012a} have shown that if gas
accumulates at some location in the disk, such as a sharp jump in viscous stress or resistivity, the gas
``bump'' can become Rossby-unstable and produce large anti-cyclonic
vortices. Initially, a number of small vortices are generated in these models that eventually merge to form a giant crescent-shaped vortex, or in other words, a
lopsided ring. These vortices could also form at the edge of a gap produced by a massive planet. As the planet opens up a gap in the disk and prevents (or at least hampers) the gas flowing toward the star, the gas accumulates at the outer edge of the gap, becomes unstable, and produces vortices \citep{Koller2003,Li2005,DeVal2007,Lin2011a,Lin2012a}. However, it is less clear how long these vortices can survive \citep{Meheut2012a}.

In addition, a massive enough planet can produce an eccentric gap and thus generate another kind of large asymmetry. \cite{Kley2006} showed that a planet with a mass larger than $3M_{\mathrm{Jup}}$ in a disk with a viscosity $\alpha\approx4\times10^{-3}$ is able to produce an eccentric disk which means that the gas parcels in the disk are on elliptic orbits. Because of Kepler's laws, the gas velocities at the apocenter of the elliptic orbits are lower than at the pericenter. As a result of this ``traffic jam'', the density $\rho$ is expected to be higher at
apocenter than at pericenter for a nearly constant mass flux ($\rho u_{\phi}$) along the ellipse. This also leads to a lopsided disk structure.  

The question arising here is how the basic differences between two models, vortex and eccentric disk, can affect the observational features of transition disks. One important difference between the eccentric structure and vortex is the density
contrast. A higher contrast is achievable if the planet produces a more eccentric disk, in
which case the eccentricity of the orbit presumably becomes apparent in the
image. Another notable difference is that the eccentric traffic jam is not comoving with the fluid, but instead is nearly stationary in the inertial frame \citep{Kley2006}. In simple words, similar to galactic spiral arms, the mass flows in from one side and flows out at the other side. As a result, this kind of asymmetry cannot trap particles, and the dust asymmetry would be similar to the gas. This has been confirmed by \cite{Hsieh2012}, who showed that in an eccentric disk with a $\sim 5M_{\mathrm{Jup}}$ planet, the dust density follows the gas density pattern. Instead, anticyclonic vortices, which are local pressure ``bumps'', act as dust traps \citep[e.g][]{Barge1995,Klahr1997,Lyra2009a}. The dust trapping may lead to an even stronger azimuthal asymmetry
in the dust continuum \citep{Birnstiel2013a}.

The objective of this paper is to study the observational appearance of asymmetries caused by a planet and answer the question when we observe a transition disk and see a large-scale asymmetric ring, whether this is due to a vortex or an eccentric disk. We aim to find out under which conditions eccentricities or vortices are the dominant sources of azimuthal asymmetries.

\section{Method and simulations} \label{method}
Both a vortex and an eccentric disk, which cause an asymmetry in a gas disk, can be produced by a massive planet and are affected by the physical parameters of the disk in the same manner, although they have different formation mechanisms. In Sec.~\ref{theo} we briefly diskuss the formation of these asymmetries and the parameters that can affect them, then we describe our numerical methods.

\subsection{Vortex and eccentric shape} 
\label{theo}
A large-scale vortex can be created by a Rossby-wave instability (RWI) and live long under suitable conditions. A RWI can be considered as the rotational analogy of the Kelvin-Helmholtz instability \citep{Lovelace1999}. \cite{Li2000} studied the condition of RWI formation in a disk with either a density bump or a density jump and concluded that a 10-20\% surface density change over a radial length scale of about the thickness of the disk is enough to perturb the rotational velocity of the gas and produce the RWI. Wherever this condition is satisfied in the disk, the instability occurs. \cite{DeVal2007} showed that not only planets as massive as Neptune and Jupiter are able to excite the RWI and generate vortices on both edges of the gap, but the vortices can also live longer than 100 orbits. A 3D study by \cite{Meheut2012a} showed that RWI can survive for long timescales ($\sim$ hundred orbits), and the authors explained that, as long as the overdensity is sustained permanently by an external driving force such as (in our case) a planet, the vortex does not decay and is not destroyed by mechanisms such as elliptical instability. Therefore, parameters such as viscosity and planet mass, which control the accretion rate through the gap and consequently the density gradient in the outer edge of the gap, can influence the formation and the lifetime of the vortex.

An eccentric gap is another by-product of gap formation by a massive planet and is affected by parameters similar to those that influence a vortex. \cite{Kley2006} investigated the response of disk eccentricity by changing planet mass and physical parameters of the disk such as viscosity and temperature. They showed that if the planet is massive enough to clear a gap that reaches the outer 1:2 Lindblad resonance, the disk becomes eccentric. Like vortex formation, the eccentric structure is also altered by viscosity and planet mass. High viscosity and lower planet mass narrow the gap and reduce the eccentricity of the disk. We ran 132 hydrodynamical simulations to study how these parameters affect the formation and lifetime of a vortex and/or an eccentric feature.

\subsection{Hydro-simulations} 
\label{numer}
We used the locally isothermal version of the FARGO code \citep{Masset2000}. Our basic model is a 2D viscous disk with a massive planet that opens a gap. We altered viscosity and mass of the planet to see how they affect the asymmetric features. We performed 12 runs in inertial frame for 1000 orbits. To distinguish between a vortex and an eccentric shape, we followed the disk surface density evolution at a high time-resolution. To do this, we conducted an extra set of ten runs per model to extend every output in each simulation for another two orbits, which saved 100 outputs during each orbit.

We considered a flared disk with an aspect ratio of $h=0.05(r/r_p)^{0.25}$. Considering the aspect ratio definition  $h(r)=H/r=c_{s}/v_{K}$, where $c_{s}$ is the local sound speed and $v_{K}$ is the Keplerian velocity, the temperature profile scales as $T \propto r^{-1/2}$. The surface density follows the relation $\Sigma=\Sigma_0~(r/r_p)^{-1}$, with $\Sigma_0=2\times10^{-4}~M_{\star}/r_p^2$, and to keep the disk viscously stable, we used a viscosity of $\nu=\alpha c_{s} H$ \citep{Shakura1973}. The viscosity parameter $\alpha$ had the values $10^{-2},\ 10^{-3}$, and $10^{-4}$ in our models. The disk extended from $r_{\mathrm{min}}=0.1 r_{p}$ to $r_{\mathrm{max}}=7.0 r_{p}$, where $r_{p}$ is the orbital radius of the planet and was used as length scale. The disk is covered by $N_{r} \times N_{s}=512 \times 757$ grid cells with logarithmic radial spacing. We chose this resolution to avoid numerical problems caused by the high density gradient at the gap edge and to have squared cells. We used the FARGO non-reflecting boundary condition to reduce the effect of wave reflections from the boundaries.

The planet mass in our simulations had the values $M_{p}=5 , 10, 15,$ and $20$ $M_{\mathrm{Jup}}$ with $M_{\mathrm{Jup}}=10^{-3} M_{\star}$. The planet was held at $r_{p}=1$ in a circular orbit. The potential of the planet $\phi$ was softened by the parameter $\epsilon=0.6R_{H}$ to avoid a singularity:

\begin{eqnarray}
	\label{eq1}
	\phi=-\frac{GM_{p}}{(r^2+\epsilon ^{2})},
\end{eqnarray}
where $R_{H}$ is the planet Hill radius of the planet and $G$ is the gravitational constant.

To apply our results to the observed features of the transition disks, we need the planet orbital radius and mass of the star in physical units. The remaining quantities can be calculated using these two numbers. In this work we considered a planet that orbits a solar-mass star at a distance of $r_{p}=20$AU. Therefore, the disk extends from $r_{\mathrm{min}}=2$AU to $r_{\mathrm{max}}=140$AU and has a mass of $M_{\mathrm{disk}}~=0.008M_{\mathrm{\odot}}$. The initial surface density $\Sigma_0$ is therefore $\Sigma_0\sim 4.44~(r/20\mathrm{AU})^{-1} \mathrm{g}\ \mathrm{cm}^{-2}$.

\begin{figure}
	\begin{center}
	\includegraphics[width=9cm]{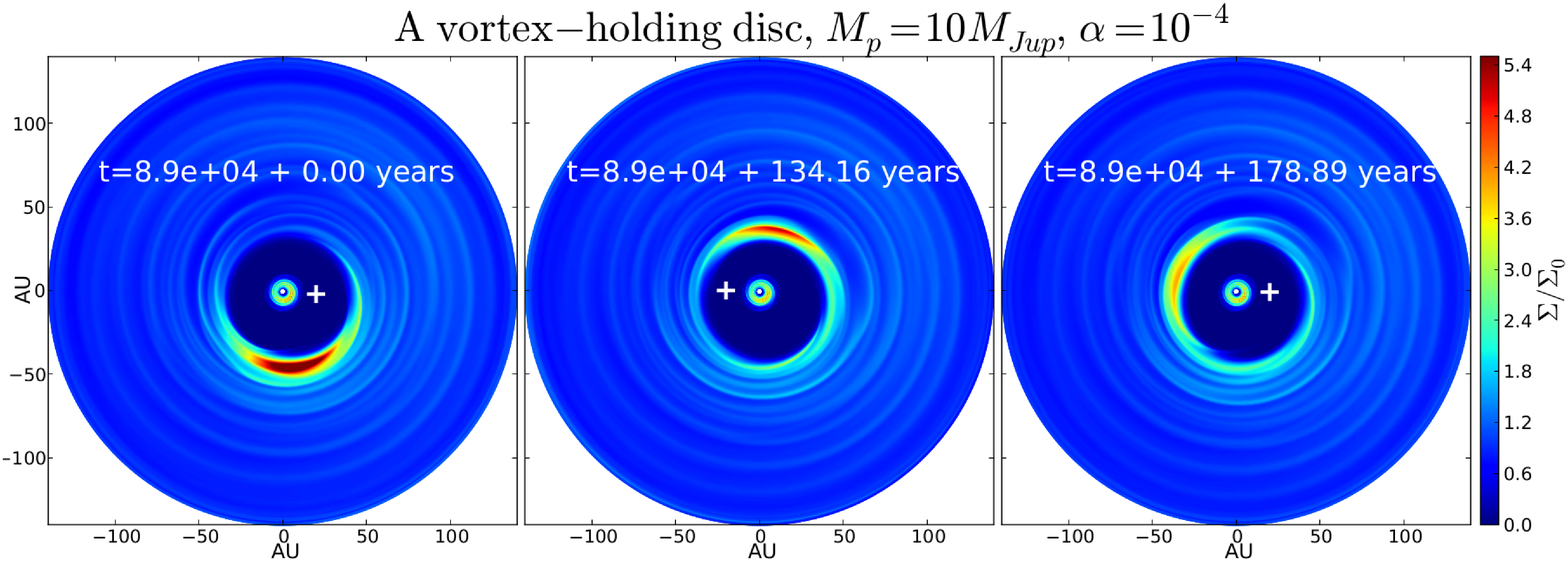}
	\includegraphics[width=9cm]{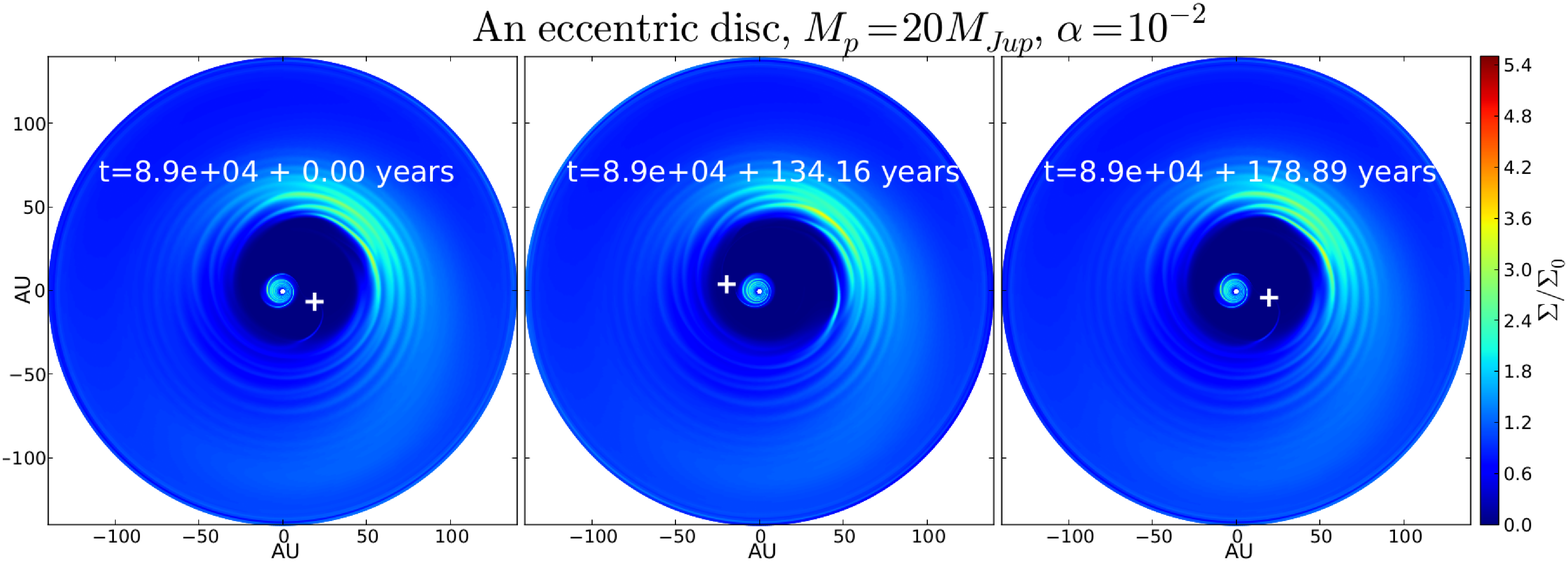}
		\caption{Evolution of a vortex-holding disk and an eccentric disk from 1000 orbits (left column) to 1002 orbits (right column). The top panels show how a vortex moves during two orbits of the planet. The bottom panels show the same snapshots as the top panels, but for an eccentric disk. Clearly, the eccentric feature is fixed during the two planetary orbits. The crosses display the planet position.}
	\label{rec}
	\end{center}
\end{figure}

\subsection{Recognizing a vortex from an eccentric shape} 
\label{recVE}
We distinguished a vortex from an eccentric asymmetry in our hydro-simulations by checking whether the structure was fixed or co-moving with the gas in the inertial frame. While a vortex revolves around the star with the local orbital frequency, an eccentric feature is almost fixed with small precession \citep{Kley2006}. Accordingly, we followed the structures in all models to decide whether the asymmetry was a vortex or an eccentric feature (Fig.~\ref{rec}).

\subsection{Dust simulations}
\label{dustsim}
To compare the dust distribution between a vortex and an eccentric feature, we conducted dust simulations for the models in Fig.~\ref{rec}. We used the numerical code developed by \cite{Zsom2011}, which solves the restricted three-body problem for dust particles in 2D spherical coordinates, taking into account gas drag that results in radial drift \citep[for details see][]{Paardekooper2007,Zsom2011}. To mimic the turbulence mixing, we allowed the particles to pace every time step ($\Delta t$) in a random direction with a step size of $l= \sqrt{D_{p}\Delta t}$, where $D_{p}=\nu/(1+{\tau_{s}}^{2})$ is the diffusion coefficient for the particles with dimensionless stopping time $\tau_{s}$ \citep[see][]{Youdin2007}. The viscosity $\nu$ was taken to be self-consistent with our hydrodynamical simulations. Because we are only interested in the dust dynamics, we did not consider the dust coagulation and fragmentation. We ran two sets of simulations for two particle masses of $0.05$ and $0.5 \mathrm{g}$. These masses correspond to sizes $\sim2$ and $\sim5 \mathrm{mm}$ in our simulations which suffer the highest and moderate radial drift at $\sim50AU$. We distributed 10000 dust particles with the same density profile as the gas at 1000 orbits and allowed particles to evolve for 200 orbits. 

\section{Results}
\begin{figure}
	\begin{center}
		\includegraphics[width=8cm]{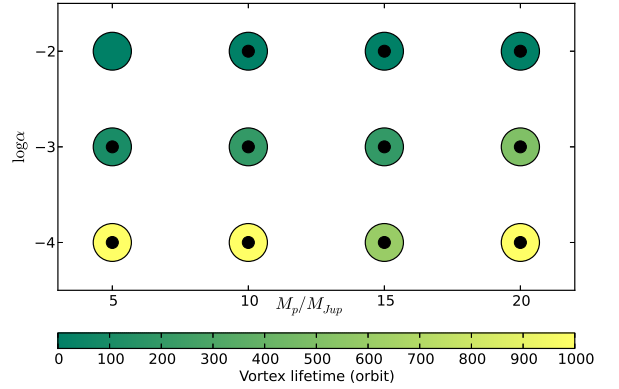}
		\caption{Dependency of eccentricity and vortex lifetimes on planet mass and viscosity. The color represents the lifetime of vortices, and the eccentric disks are marked by a dot. For $M_{p}=15M_{\mathrm{Jup}}$ and $\alpha=10^{-4}$, the asymmetries are extremely mixed and indistinguishable after 600 orbits.}
		\label{cool}
	\end{center}
\end{figure}

Figure~\ref{cool} summarizes of the results of our models in terms of the presence or absence of a vortex and/or an eccentricity. The horizontal and vertical axes are the mass of the planet and the logarithm of the $\alpha$ viscous parameter. According to Fig.~\ref{cool}, an eccentric shape is a very common feature in our models. This figure shows that a vortex can be created and survive until the end of the simulations for a low viscosity, i.e., $\alpha~=~10^{-4}$. 

In none of the simulations with the highest viscosity parameter $\alpha=10^{-2}$, was a vortex formed. In the model with a $5 M_{\mathrm{Jup}}$ planet, a narrow circular gap was created and stayed until the end of the simulations. A more massive planet $M_{p}=10M_{\mathrm{Jup}}$ starts to form an eccentric gap from 200 orbits and establishes this after 400 orbits ($\sim 3.6 \times 10^{4}$ years). For more massive planets, the disk becomes eccentric even at 200 orbits ($\sim 1.8 \times 10^{4}$ years). 

For an intermediate value of the viscosity, i.e., $\alpha=10^{-3}$, the disk passed through three stages. First a vortex appeared in the disk and survived until 200-300 orbits. The lifetime of the vortex greatly depends on mass of the planet. The more massive the planet, the longer the vortex survives. During the second stage, which lasts until orbit 800 for $M_{p}=5M_{\mathrm{Jup}}$ and until orbit 500 for the remaining masses, the features are mostly indistinguishable and mixed. Therefore, even if we observe a moving denser region or a vague asymmetry, we consider it to be neither a vortex nor an eccentric disk. Eventually, the planet succeeds in reforming the gap into an eccentric shape and the disk keeps this until the end of the simulation.

The models with a low eccentricity of $\alpha=10^{-4}$ produce the most interesting results for the vortex formation. For lower planet masses, i.e., $M_{p}=5M_{\mathrm{Jup}}$ and $10 M_{\mathrm{Jup}}$, the vortex is the dominant feature. Although the disk is slightly eccentric, the asymmetric density feature caused by it is not strong enough to be distinguished, and therefore the vortex remains the strongest asymmetry in the disk. On the other hand, the eccentric feature in the models with more massive planets $M_{p}=15M_{\mathrm{Jup}}$ and $20 M_{\mathrm{Jup}}$ is strong to such a degree that it can mix with the vortex. When the two features overlap, the density contrast is the the highest. 

\begin{figure}
	\begin{center}
		\includegraphics[width=9cm]{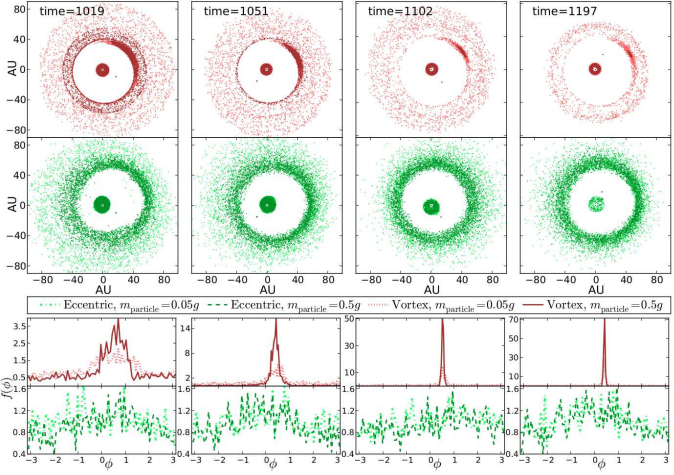}
		\caption{Dust distribution comparison at 1019, 1051, 1102, and 1197 orbits. We chose these snapshots because the vortex is located at the same azimuthal position of the eccentric feature which facilitates the comparison of the azimuthal particle concentration. The top and middle rows reflect the difference between the dust trapping in a vortex-holding disk and an eccentric disk. The light and dark colors represent the $0.05\mathrm{g}$ and $0.5\mathrm{g}$ particles. In the bottom panels, we plot the dust enhancement factor against azimuth at each corresponding snapshot.}
		\label{dust}
	\end{center}
\end{figure}

Figure \ref{dust} illustrates the results of the dust simulations for the two disk models in Fig.~\ref{rec}. The first row of Fig.~\ref{dust} shows the dust distribution for the disk with the vortex. In this case, particles move toward the vortex, and after 200 orbits most of the dust is accumulated in the vortex center. The middle panels of Fig.~\ref{dust} display the the dust distribution in the eccentric case. In contrast to the vortex case, the particles are only trapped in the radial direction, which produces a ring that is wide because of the high diffusion ($\alpha=10^{-2}$). In the top panels the turbulence stirring is not efficient because of the low viscosity and the results would be very similar to those of the models without stirring. However, the particles in middle panels are less concentrated around the outer edge of the gap because of the higher diffusion coefficient. After the semi-steady state is reached, no strong azimuthal asymmetry is produced. We define the \emph{dust enhancement factor} $f$ as

\begin{eqnarray}
	\label{eq2}
	f(\phi)=\frac{(M_{\mathrm{dust}}/M_{\mathrm{gas}})_{t}}{(M_{\mathrm{dust}}/M_{\mathrm{gas}})_{t=1000\ \mathrm{orbits}}},
\end{eqnarray}

\noindent which is calculated for $r\in [30,140]\mathrm{AU}$ and the azimuthal slice between $\phi$ and $\phi+\Delta \phi$. We plot $f(\phi)$ in the bottom panels of Fig.~\ref{dust} to compare the dust azimuthal accumulation of our models. These plots show that the dust enhancement factor in a vortex is much higher than the value for the eccentric case. This means that, the vortex produces a stronger azimuthal asymmetry in a narrower azimuthal range than the eccentric disk.

\section{Observational consequences}

The difference between azimuthally density contrast in a vortex-holding and an eccentric disk is a very useful tool for recognizing them observationally. Our results show that we need $\sim 50 \times (R_{s}/20AU)^{1.5}$ years ($R_{s}$ is the asymmetry position) to distinguish an eccentric feature from a vortex using the comparison of the position of the azimuthal density variation. However, different observational features at (sub-)millimeter wavelengths are fortunately diskernible in one single snapshot. When a planet carves out a gap in a disk, particle trapping in the \emph{radial} direction would happen for both cases \citep[see e.g][]{Pinilla2012b}, while particle trapping in \emph{azimuthal} direction can only happen in the vortex case. Because a traffic jam in an eccentric disk is no particle trap, the azimuthal contrast in the continuum would be as high as the contrast in the gas surface density, which can reach values of $\Sigma/ \Sigma_0\sim$~3 (Fig.~\ref{rec}), and the dust azimuthal extension of the asymmetry would be similar to gas. Conversely, vortices are indeed particle traps and can be long-lived (up to $10^{5}$ years) when very massive planets (15-20~$M_{\mathrm{Jup}}$) and moderate viscosity ($\alpha=10^{-4}$) are able to maintain the density bumps and the RWI. Thus, the vortices can create strong  variations in the azimuthal dust distribution of millimeter particles and the asymmetry feature would be less azimuthally extended for the dust than for the gas \citep{Birnstiel2013a}. Hence, stronger and more compact dust azimuthal asymmetry is expected for a vortex than for an eccentric disk. The results shown in Fig.~\ref{dust} give us a good impression of what is expected with millimeter observations with high sensitivity and angular resolution. It is important to notice that neither in a vortex nor in an eccentric case, we expect the location of the planet along its orbit to be correlated with the location of the vortex or ellipse of disks.

\section{Conclusion}

One exciting explanation of the wide gaps observed in transition disks is the interaction of a massive planet with the disk. When the planet is massive enough and moderate values for the disk viscosity are assumed, two different scenarios can create azimuthal variations in the gas surface density: eccentric disks and a vortex excited by the RWI. The azimuthal contrast of the gas surface density varies for both cases and is much higher for the vortex situation. In this case, particle trapping is possible and very high azimuthal variations are expected at millimeter wavelengths, as is currently observed with ALMA.

\begin{acknowledgements}
We thank \emph{T.~Birnstiel, M.~Benisty, A. Juhasz, E.~van Dishoeck, N.~van der Marel, Z.~Regaly, and H.~Meheut} for the fruitful discussions. S.~Ataiee and P.~Pinilla acknowledge the CPU time for running simulations in bwGRiD, member of the German D-Grid initiative, funded by the Ministry for Education and Research (Bundesministerium f\"ur Bildung und Forschung) and the Ministry for Science, Research and Arts Baden-Wuerttemberg (Ministerium f\"ur Wissenschaft, Forschung und Kunst Baden-W\"urttemberg). This work was financed partly by a scholarship of the Ministry of Science, Research, and Technology of Iran, and partly by the third funding line of the German Excellence Initiative.
\end{acknowledgements}

\bibliographystyle{aa} 
\bibliography{lib.bib}

\begin{thebibliography}{34}
\expandafter\ifx\csname natexlab\endcsname\relax\def\natexlab#1{#1}\fi

\bibitem[{Andrews {et~al.}(2011)Andrews, Wilner, Espaillat, Hughes, Dullemond,
  McClure, Qi, \& Brown}]{Andrews2011}
Andrews, S.~M., Wilner, D.~J., Espaillat, C., {et~al.} 2011, ApJ, 732, 42

\bibitem[{Barge \& Sommeria(1995)}]{Barge1995}
Barge, P. \& Sommeria, J. 1995, A\&A, 295, L1

\bibitem[{Birnstiel {et~al.}(2013)Birnstiel, Dullemond, \&
  Pinilla}]{Birnstiel2013a}
Birnstiel, T., Dullemond, C.~P., \& Pinilla, P. 2013, A\&A, 550, L8

\bibitem[{Brown {et~al.}(2009)Brown, Blake, Qi, Dullemond, Wilner, \&
  Williams}]{Brown2009}
Brown, J.~M., Blake, G.~a., Qi, C., {et~al.} 2009, ApJ, 704, 496

\bibitem[{Casassus {et~al.}(2013)Casassus, van~der Plas, M, Dent, Fomalont,
  Hagelberg, Hales, Jord\'{a}n, Mawet, M\'{e}nard, Wootten, Wilner, Hughes,
  Schreiber, Girard, Ercolano, Canovas, Rom\'{a}n, \& Salinas}]{Casassus2013}
Casassus, S., van~der Plas, G., M, S.~P., {et~al.} 2013, Nature, 493, 191

\bibitem[{D'Alessio {et~al.}(1998)D'Alessio, Canto, Calvet, \&
  Lizano}]{D'AlessioPaola1998}
D'Alessio, P., Canto, J., Calvet, N., \& Lizano, S. 1998, ApJ, 500, 411

\bibitem[{De~Val-Borro {et~al.}(2007)De~Val-Borro, Artymowicz, D'Angelo, \&
  Peplinski}]{DeVal2007}
De~Val-Borro, M., Artymowicz, P., D'Angelo, G., \& Peplinski, A. 2007, A\&A,
  471, 1043

\bibitem[{Fukagawa {et~al.}(2004)Fukagawa, Hayashi, Tamura, Itoh, Hayashi,
  Oasa, Takeuchi, Morino, Murakawa, {Oya, S. and Yamashita, T. and Suto},
  Mayama, Naoi, Ishii, , Pyo, Nishikawa, Takato, Usuda, Ando, Iye, Miyama, \&
  Kaifu}]{Fukagawa2004}
Fukagawa, M., Hayashi, M., Tamura, M., {et~al.} 2004, ApJ, 605, L53

\bibitem[{Hsieh \& Gu(2012)}]{Hsieh2012}
Hsieh, H.-F. \& Gu, P.-G. 2012, ApJ, 760, 119

\bibitem[{Klahr \& Henning(1997)}]{Klahr1997}
Klahr, H.~H. \& Henning, T. 1997, Icarus, 229, 213

\bibitem[{Kley \& Dirksen(2006)}]{Kley2006}
Kley, W. \& Dirksen, G. 2006, A\&A, 377, 369

\bibitem[{Koller {et~al.}(2003)Koller, Li, \& Lin}]{Koller2003}
Koller, J., Li, H., \& Lin, D. 2003, ApJ, 596, L91

\bibitem[{Li {et~al.}(2000)Li, Finn, \& Lovelace}]{Li2000}
Li, H., Finn, J., \& Lovelace, R. 2000, ApJ, 533, 1023

\bibitem[{Li {et~al.}(2005)Li, Li, Koller, Wendroff, Liska, Orban, Liang, \&
  Lin}]{Li2005}
Li, H., Li, S., Koller, J., {et~al.} 2005, ApJ, 624, 1003

\bibitem[{Lin(2012)}]{Lin2012a}
Lin, M.-K. 2012, MNRAS, 426, 3211

\bibitem[{Lin \& Papaloizou(2011)}]{Lin2011a}
Lin, M.-K. \& Papaloizou, J. C.~B. 2011, MNRAS, 415, 1445

\bibitem[{Lovelace {et~al.}(1999)Lovelace, Li, Colgate, \&
  Nelson}]{Lovelace1999}
Lovelace, R. V.~E., Li, H., Colgate, S.~a., \& Nelson, a.~F. 1999, ApJ, 513,
  805

\bibitem[{Lyra {et~al.}(2009)Lyra, Johansen, Klahr, \& Piskunov}]{Lyra2009a}
Lyra, W., Johansen, A., Klahr, H., \& Piskunov, N. 2009, A\&A, 493, 1125

\bibitem[{Lyra \& {Mac Low}(2012)}]{Lyra2012a}
Lyra, W. \& {Mac Low}, M.-M. 2012, ApJ, 756, 62

\bibitem[{Masset(2000)}]{Masset2000}
Masset, F. 2000, A\&AS, 141, 165

\bibitem[{Meheut {et~al.}(2012)Meheut, Keppens, Casse, \& Benz}]{Meheut2012a}
Meheut, H., Keppens, R., Casse, F., \& Benz, W. 2012, A\&A, 542, A9

\bibitem[{Min {et~al.}(2011)Min, Dullemond, Kama, \& Dominik}]{Min2011}
Min, M., Dullemond, C., Kama, M., \& Dominik, C. 2011, Icarus, 212, 416

\bibitem[{Muto {et~al.}(2012)Muto, Grady, Hashimoto, Fukagawa, Hornbeck, Sitko,
  Russell, Werren, Cur\'{e}, Currie, Ohashi, Okamoto, Momose, Honda, Inutsuka,
  Takeuchi, Dong, Abe, Brandner, Brandt, Carson, Egner, Feldt, Fukue, Goto,
  Guyon, Hayano, Hayashi, Hayashi, Henning, Hodapp, Ishii, Iye, Janson,
  Kandori, Knapp, Kudo, Kusakabe, Kuzuhara, Matsuo, Mayama, McElwain, Miyama,
  Morino, Moro-Martin, Nishimura, Pyo, Serabyn, Suto, Suzuki, Takami, Takato,
  Terada, Thalmann, Tomono, Turner, Watanabe, Wisniewski, Yamada, Takami,
  Usuda, \& Tamura}]{Muto2012}
Muto, T., Grady, C.~a., Hashimoto, J., {et~al.} 2012, ApJ, 748, L22

\bibitem[{Oppenheimer {et~al.}(2008)Oppenheimer, Brenner, Hinkley, Zimmerman,
  Sivaramakrishnan, Soummer, Kuhn, Graham, Perrin, Lloyd, Roberts, \&
  Harrington}]{Oppenheimer2008}
Oppenheimer, B.~R., Brenner, D., Hinkley, S., {et~al.} 2008, ApJ, 679, 1574

\bibitem[{Paardekooper(2007)}]{Paardekooper2007}
Paardekooper, S. 2007, A\&A, 462, 355

\bibitem[{Pi\'{e}tu {et~al.}(2005)Pi\'{e}tu, Guilloteau, \& Dutrey}]{Pietu2005}
Pi\'{e}tu, V., Guilloteau, S., \& Dutrey, A. 2005, A\&A, 443, 945

\bibitem[{Pinilla {et~al.}(2012)Pinilla, Benisty, \& Birnstiel}]{Pinilla2012b}
Pinilla, P., Benisty, M., \& Birnstiel, T. 2012, A\&A, 545, A81

\bibitem[{Pinte {et~al.}(2006)Pinte, Menard, Duchene, \& Bastien}]{Pinte2006}
Pinte, C., Menard, F., Duchene, G., \& Bastien, P. 2006, A\&A, 459, 797

\bibitem[{Rameau {et~al.}(2012)Rameau, Chauvin, a.~M.~Lagrange, Th\'{e}bault,
  Milli, Girard, \& Bonnefoy}]{Rameau2012}
Rameau, J., Chauvin, G., a.~M.~Lagrange, {et~al.} 2012, A\&A, 546, A24

\bibitem[{Reg\'{a}ly {et~al.}(2012)Reg\'{a}ly, Juh\'{a}sz, S\'{a}ndor, \&
  Dullemond}]{Regaly2011}
Reg\'{a}ly, Z., Juh\'{a}sz, A., S\'{a}ndor, Z., \& Dullemond, C.~P. 2012,
  MNRAS, 419, 1701

\bibitem[{Shakura \& Sunyaev(1973)}]{Shakura1973}
Shakura, N. \& Sunyaev, R. 1973, A\&A, 24, 337

\bibitem[{Woitke {et~al.}(2009)Woitke, Kamp, \& Thi}]{Woitke2009}
Woitke, P., Kamp, I., \& Thi, W.-F. 2009, A\&A, 501, 383

\bibitem[{Youdin \& Lithwick(2007)}]{Youdin2007}
Youdin, A.~N. \& Lithwick, Y. 2007, Icarus, 192, 588

\bibitem[{Zsom {et~al.}(2011)Zsom, S\'{a}ndor, \& Dullemond}]{Zsom2011}
Zsom, a., S\'{a}ndor, Z., \& Dullemond, C.~P. 2011, A\&A, 527, A10

\end{thebibliography}

\end{document}